\documentclass[a4paper,twoside,10pt]{letter}
\usepackage{graphicx,saj,multicol,subeqnarray}

\def\papertype{\ \hfill\ Original Scientific Paper}

\def\kms{$\mathrm{km\;s}^{-1}$}

\def\Msun{{\rm\,M_\odot}}

\def\h3{$h_{3}$}
\def\h4{$h_{4}$}

\begin{document}
\baselineskip=3.1truemm
\columnsep=.5truecm
\newenvironment{lefteqnarray}{\arraycolsep=0pt\begin{eqnarray}}
{\end{eqnarray}\protect\aftergroup\ignorespaces}
\newenvironment{lefteqnarray*}{\arraycolsep=0pt\begin{eqnarray*}}
{\end{eqnarray*}\protect\aftergroup\ignorespaces}
\newenvironment{leftsubeqnarray}{\arraycolsep=0pt\begin{subeqnarray}}
{\end{subeqnarray}\protect\aftergroup\ignorespaces}
%


\markboth{\eightrm THE TOTAL MASS OF THE EARLY-TYPE GALAXY
NGC~4649 (M60)}
{\eightrm S. SAMUROVI{\' C} {\lowercase{and}} M.M. \'CIRKOVI\'C}

{\ }

\publ

\type

{\ }


\title{THE TOTAL MASS OF THE EARLY-TYPE GALAXY
NGC~4649 (M60)}


\authors{S. Samurovi{\' c} and M. M. \'Cirkovi\'c}

\vskip3mm


\address{Astronomical Observatory, Volgina 7, 11060 Belgrade 38, Serbia}

\centerline{E-mail: {\it srdjan@aob.bg.ac.yu}}

\centerline{E-mail: {\it mcirkovic@aob.bg.ac.yu}}


\dates{July 21, 2008}{August 21, 2008}


\summary{In this paper we analyze the problem of the total mass
and the total mass-to-light ratio of the early-type galaxy
NGC~4649 (M60). We have used two independent techniques: the
X-ray methodology which is based on the temperature of the X-ray
halo of NGC~4649 and the tracer mass estimator (TME) which uses
globular clusters (GCs) observed in this galaxy. We calculated
the mass in Newtonian and MOdified Newtonian Dynamics (MOND)
approaches and found that interior to 3 effective radii ($R_e$)
there is no need for large amounts of dark matter. Beyond $3R_e$
dark matter starts to play important dynamical role. We also
discuss possible reasons for the discrepancy between the
estimates of the total mass based on X-rays and TME in the outer
regions of NGC~4649.}



\begin{multicols}{2}

\section{1. INTRODUCTION}

The problem of dark matter in  early--type  galaxies is still one
of the most important unsolved (and, we might add, poorly
understood) problems in the contemporary extragalactic astronomy
and cosmology (e.g.~Samurovi\'c 2007, Chap.~1,  for a detailed
introduction). For decades it was almost universally considered that the
giant ellipticals are paramount reservoirs of dark matter, a
prejudice partially based on the poor understanding of the
Faber-Jackson relation linking the central velocity dispersion and
the galactic luminosity. The general picture which seems to emerge
with contemporary observational data, suggests that interior to
$\sim$2--3 effective radii, $R_e$, (inner regions) dark matter
does not play any important dynamical role (e.g.~Samurovi\'c and
Danziger 2005). At larger galactocentric distances (beyond $\sim
3-4R_e$) it starts to play more important role;  its
contribution, however, varies from case to case (e.g.~Samurovi\'c
and Danziger 2006, Samurovi\'c 2006). Overall, there is still
considerably smaller amount of observational data for elliptical
than for spiral galaxies, which makes establishing general
results very difficult.

In order to find a typical mass-to-light ratio characteristic for 
early-type galaxies, one can use the paper by van der Marel (1991) who found on 
a sample of 37 bright ellipticals that this quantity in the $B$-band is: 
$M/L_B=(5.95\pm 0.25)h_{50}$; thus $M/L_B=8.33\pm 0.35$ for $h=0.70$ (the value 
of the Hubble constant used in this paper; see Tegmark et al.\ 2004). This 
sample pertained to inner parts of ellipticals, and therefore we  consider the 
absolute upper limit for the visible (stellar) component $M/L_B\sim 9-10$. Any 
inferred mass-to-light ratio above $\sim 10$ in a given region would thus 
imply the existence of unseen (dark) matter there. Note that this conclusion 
is independent on whether the inner regions studied by van der Marel (1991) 
also contain significant amounts of dark matter (which is highly unlikely). We 
stress that the acceptance of large quantities of dark matter in the 
astrophysical community is not unanimous. One possible (and in some cases 
seemingly viable) alternative is MOND (MOdified Newtonian Dynamics), pionereed 
by Milgrom (1983), which will also be tested in the present paper. 

Recently, new kinematical data of NGC~4649 which extend out to $\sim 6R_e$, 
based on the observations of globular clusters (GCs), became available (Lee et 
al.~2008). Samurovi\'c and \'Cirkovi\'c (2008, hereafter Paper I) used these 
data to study the velocity dispersion profile using the Jeans equation and 
found that the mass-to-light ratio $M/L_B\sim 7$ provides a satisfactory fit 
of the velocity dispersion throughout the  whole galaxy which implies the 
scarcity of dark matter in this galaxy. In the present paper our aim is to 
employ other independent techiques to calculate the total mass (and 
mass-to-light ratio) and compare the estimates obtained using different 
approaches (both Newtonian and MOND). 

The plan of the paper is as follows: In Section~2 we present the basic 
observational data related to NGC~4649 which we use in this paper. In 
Section~3 we calculate the total mass of NGC~4649 in Newtonian 
mass-follows-light approach based on two different techiques: X-rays and 
tracer mass estimator (TME). In this section we also apply the virial theorem 
in order to calculate the total mass of this galaxy. In Section~4 we calculate 
the mass of NGC~4649 in MOND approach (using three  different relationships) 
again based on the two afforementioned techiques, X-rays and TME. Finally, in 
Section~5 we discuss our findings and present the conclusions. 

\section{2. THE BASIC DATA}

NGC~4649 (M60) is a giant elliptical galaxy in the Virgo cluster. It has a 
nearby companion, NGC~4647 (Sc galaxy at $2.'5$ from the center of NGC~4649). 
The systemic velocity of NGC~4649 is $v_{\hbox{\eightrm{\rm hel}}}=1117\pm 6$ 
\kms.  Hereafter we use the distance based on this systemic 
velocity -- for the recent WMAP estimate of the Hubble constant (Tegmark et 
al. 2004; Komatsu et al.~2008), $h_0=0.70$ we obtain $d=15.96$ Mpc;  in this 
case 1 arcsec corresponds to 77.5 pc. For the effective radius, as in Paper I, 
we take $R_e=90$ arcsec (which is equal to  $6.97$ kpc for the distance which 
we use), the value taken from the paper by  Kim et al.~(2006). Note that there 
are different estimates in the literature, for example, Lee et al.~(2008) took 
$R_e=110$ arcsec and according to the RC3 catalog (de Vaucouleurs et 
al.~1991), $R_e=69$ arcsec. Our adopted value is therefore an intermediate 
one. The discrepancy does not significantly impact our conclusions. 

The kinematics of NGC~4649 which includes velocity dispersion and symmetric 
and asymmetric departures from the Gaussian is given  in Fig. 1 of Paper I. 
The radial distribution of GCs is given  in Fig. 2 of Paper I: it was shown 
that beyond $\sim  1$ arcmin the  surface number density can be fitted with a 
power law: $\Sigma \propto r^{-\gamma}$, where $\gamma=1.285$ and $r$ is the 
distance to the galaxy center in projection. 

As in Paper I we here use the observations of GCs from the study of Lee et 
al.~(2008). This sample  consists of 121 GCs (83 blue and 38 red  GCs) and in 
all our calculations we always deal with the total sample in order to have a 
larger number of GCs per each bin. 

\section{\bf 3. TOTAL MASS IN NEWTONIAN GRAVITY}

In this Section we present the total mass of NGC~4649 (and the total 
mass-to-light ratio in the $B$-band) in Newtonian gravity and in next Section 
we calculate the same quantities in MOND gravity. 

To calculate Newtonian mass we use two different, independent techniques: 
X-rays and tracer mass estimator (TME). The estimates and comparison are given 
in Fig. 1  and in Table~1. In this Section we also use scalar virial theorem  
as a check for the estimate of the total mass of NGC~4649. 

\subsection{\bf 3.1 X-RAY ESTIMATE}

The total mass of NGC~4649 interior to the radius $r$ based
on the X-ray observations is calculated using the following
equation taken from Kim and Fabbiano (1995):
\begin{eqnarray}
M_T&=&1.8\times 10^{12}(3\beta +\alpha)
\left ( {T\over {\hbox{\eightrm{\rm 1 keV}}}} \right )
\left ( {r\over 1000{}^{\prime\prime} }\right ) \cr
& & \left ({d\over  {\hbox{\eightrm{\rm 10 Mpc}}}}\right ) \Msun,
\label{eqn:TOT1}
\end{eqnarray}
where the exponent $\alpha$ is related to the temperature of the X-ray halo 
($T\sim r^{-\alpha}$) and in our calculations taken to be zero, while $\beta$ 
is the slope used in the analytic King approximation model and taken to be 
$\beta = 0.50$ (Brown and Bregman 2001). The mass-to-light ratio in the 
$B$--band  calculated as a function of radius $r$ is given as: 
\begin{eqnarray}
{M_T\over L_B}&=&1.16\times 10^{-2} 10^{B\over 2.5} (3\beta +\alpha)
\left ( {T\over {\hbox{\eightrm{\rm 1 keV}}}} \right ) \cr
& & \left ( {r\over
1000{}^{\prime\prime} }\right ) 
\left ({d\over  {\hbox{\eightrm{\rm 10
Mpc}}}}\right )^{-1} \!, \label{eqn:ML1}
\end{eqnarray}
where $B$ is the  B magnitude of the galaxy inside radius $r$ (Kim and Fabbiano 
1995). In our estimates for the temperature in Fig.~1 we have used the results 
of Randall et al.~(2006): $T=0.80\pm 0.05$ keV.

It is important to note that the two relations above assume the existence of 
hydrostatic equilibrium, which is almost universal in interpreting the X-ray 
profiles. Based on the recent work of Diehl and Statler (2007) the optical and 
X-ray isophotes for NGC~4649 are very close: for X-rays (based on the CHANDRA 
observations) ellipticity $\epsilon_{\hbox{\eightrm{X}}}=0.08\pm 0.03$ and 
position angle ${\hbox{\rm P.A.}}_{\hbox{\eightrm X}}=95\pm 27$ whereas in the 
optical domain $\epsilon_{\hbox{\eightrm{opt}}}=0.18\pm 0.01$ and ${\hbox{\rm 
P.A.}}_{\hbox{\eightrm opt}}= 104.0\pm 0.4$. We interpret this as an indication
of hydrostatic equilibrium; for NGC~1399 the discrepancy between these two 
quantities in X-ray and optical domain is much larger: 
$\epsilon_{\hbox{\eightrm{X}}}=0.34\pm 0.04$, 
$\epsilon_{\hbox{\eightrm{opt}}}=0.10\pm 0.1$,  ${\hbox{\rm 
P.A.}}_{\hbox{\eightrm X}}=179\pm 7$ and ${\hbox{\rm P.A.}}_{\hbox{\eightrm 
opt}}=107.4\pm 0.1$.  The estimate of the total mass of NGC~1399 based on the X-
rays is given in Samurovi\'c and Danziger (2006). 

\subsection{\bf 3.2 TRACER MASS ESTIMATOR}

Evans et al. (2003) introduced a "tracer mass estimator" method
which provides an estimate of the enclosed mass based on the
projected positions and line-of-sight velocities of a given tracer
population (such as GCs in our case). We assume that the tracer
population is spherically symmetric and has a number density that
obeys a power law:
\begin{equation}
\rho(r)=\rho_0\left ({a\over r}\right )^{\gamma_1},
\label{eqn:rho1}
\end{equation}
where $a$ is constant, and radius $r$  ranges between
$r_{\hbox{\eightrm{{in}}}}$ and  $r_{\hbox{\eightrm{out}}}$,
inner and outer points of the given population, respectively. The
parameter $\gamma_1=2.285$ (we label this parameter $\gamma_1$ to avoid 
confusion with the exponent found in Fig. 2 of Paper I) was
determined using the surface density of the tracer population
between $r_{\hbox{\eightrm{{in}}}}$ and
$r_{\hbox{\eightrm{out}}}$. Evans et al.~give the following formula for the 
mass (supported by random motion) enclosed within  $r_{\hbox{\eightrm{{in}}}}$ 
and $r_{\hbox{\eightrm{out}}}$  for the isotropic ("iso" in formulae below) 
case ("los" stands for line-of-sight): 
\begin{equation}
M_p = {C_{\hbox{\eightrm{iso}}} \over G N} \sum_i v_{\hbox{\eightrm{los}}i}^2 R_i,
\label{eqn:tme1}
\end{equation}
where $R_i$ is the projected position of the $i$-th object
relative to the center of the galaxy, $N$ is the size of the
tracer population and the constant $C_{\hbox{\eightrm\rm iso}}$ is
given as:
\begin{equation}
C_{\hbox{\eightrm{iso}}} = {4\; \gamma_1 \over \pi}
    {4\!-\!\alpha\!-\!\gamma_1\over 3\!-\!\gamma_1}
    {1\!-\!(r_{\hbox{\eightrm{{in}}}}/_{\hbox{\eightrm{out}}})^{3-\gamma_1} \over
     1\!-\!(r_{\hbox{\eightrm{{in}}}}/_{\hbox{\eightrm{out}}})^{4\!-\!\alpha\!-\!\gamma_1}}.
\label{eqn:tme1_const}
\end{equation}
This expression is for the case of an isothermal potential
(gravity field is assumed to be scale-free, $\psi=-v_0^2\log r$,
see Evans et al.~2003 for details); we have also used the
isotropic case (for visible matter; for dark matter nothing is
assumed) for the sake of simplicity, given the small departures
from zero of the parameter $s_4$ which describes asymmetric
departures from the Gaussian (see Paper I).

To obtain the total mass of the galaxy we must also take into account the 
rotational component which is equal to: 

\begin{equation}
M_{\hbox{\eightrm{rot}}}=
{v_{\hbox{\eightrm{rot}}}^2R_{\hbox{\eightrm{out}}}\over G},
\label{eqn:mrot}
\end{equation}
where $R_{\hbox{\eightrm{out}}}$ is the outermost tracer projected radius in 
the sample. Therefore, the total mass of NGC~4649 is equal to the sum: 
\begin{equation}
M_{\hbox{\eightrm{tot}}}=  M_p + M_{\hbox{\eightrm{rot}}}.
\label{eqn:mtot}
\end{equation}

\end{multicols}

\vskip.5cm
\vfill\eject
\centerline{{\bf Table 1.} Newtonian and MOND mass and M/L estimates for 
NGC~4649} 
\vskip2mm

\centerline {\it The table was split in two parts in this preprint.} 

{\begin{tabular}{ccccccc}
\hline

\noalign{\smallskip}
\multicolumn{1}{c}{r} &
\multicolumn{1}{c}{$M^{\hbox{\eightrm{xray}}}_{\hbox{\eightrm{tot}}}$} &
\multicolumn{1}{c}{$M^{\hbox{\eightrm{xray}}}/L_B$} &
\multicolumn{1}{c}{$M^{\hbox{\eightrm{TME}}}_{\hbox{\eightrm{tot}}}$} &
\multicolumn{1}{c}{$M^{\hbox{\eightrm{TME}}}/L_B$} &
\multicolumn{1}{c}{$M^{\hbox{\eightrm{M,sim}}}_{\hbox{\eightrm{tot}}}$} &
\multicolumn{1}{c}{$M^{\hbox{\eightrm{M,sim}}}/L_B$} \\
\noalign{\smallskip}
\multicolumn{1}{c}{($'$)} &
\multicolumn{1}{c}{$10^{11}M_\odot$} &
\multicolumn{1}{c}{$M_\odot/L_\odot$} &
\multicolumn{1}{c}{$10^{11}M_\odot$} &
\multicolumn{1}{c}{$M_\odot/L_\odot$} &
\multicolumn{1}{c}{$10^{11}M_\odot$} &
\multicolumn{1}{c}{$M_\odot/L_\odot$} \\
\multicolumn{1}{c}{(1)} &
\multicolumn{1}{c}{(2)} &
\multicolumn{1}{c}{(3)} &
\multicolumn{1}{c}{(4)} &
\multicolumn{1}{c}{(5)} &
\multicolumn{1}{c}{(6)} &
\multicolumn{1}{c}{(7)} \\
\hline
\noalign{\smallskip}

$<$2  &   $ 4.3 \pm 0.6$ &  $ 4.8 \pm 0.7$   &  $ 4.5 \pm 0.4 $ &  $ 4.8 \pm 0.4$   &  $ 3.8  \pm 0.5$ & $ 4.2  \pm  0.6$  	\\
$<$4  &   $ 8.6 \pm 1.2$ &  $12.0 \pm 1.7$   &  $ 8.9 \pm 1.2 $ &  $11.9 \pm 1.6$   &  $ 6.7  \pm 0.9$ & $ 9.4  \pm  1.3$      \\
$<$6  &   $12.9 \pm 1.8$ &  $20.4 \pm 2.9$   &  $12.8 \pm 2.1 $ &  $19.3 \pm 3.2$   &  $ 9.0  \pm 1.3$ & $14.3  \pm  2.0$      \\
$<$9  &   $19.3 \pm 2.8$ &  $32.8 \pm 4.6$   &  $15.2 \pm 3.0 $ &  $24.6 \pm 4.9$   &  $11.8  \pm 1.7$ & $20.0  \pm  2.8$      \\
\noalign{\smallskip}
\hline
\noalign{\medskip}
\end{tabular}}

{\begin{tabular}{ccccc}
\hline

\noalign{\smallskip}
\multicolumn{1}{c}{r} &
\multicolumn{1}{c}{$M^{\hbox{\eightrm{M,std}}}_{\hbox{\eightrm{tot}}}$} &
\multicolumn{1}{c}{$M^{\hbox{\eightrm{M,std}}}/L_B$} &
\multicolumn{1}{c}{$M^{\hbox{\eightrm{M,toy}}}_{\hbox{\eightrm{tot}}}$} &
\multicolumn{1}{c}{$M^{\hbox{\eightrm{M,toy}}}/L_B$} \\
\noalign{\smallskip}
\multicolumn{1}{c}{($'$)} &
\multicolumn{1}{c}{$10^{11}M_\odot$} &
\multicolumn{1}{c}{$M_\odot/L_\odot$} &
\multicolumn{1}{c}{$10^{11}M_\odot$} &
\multicolumn{1}{c}{$M_\odot/L_\odot$} \\
\multicolumn{1}{c}{(1)} &
\multicolumn{1}{c}{(8)} &
\multicolumn{1}{c}{(9)} &
\multicolumn{1}{c}{(10)} &
\multicolumn{1}{c}{(11)} \\
\hline
\noalign{\smallskip}

$<$2   &$ 4.2 \pm 0.6$ & $ 4.7  \pm  0.7$ &$2.9 \pm 0.4$ & $ 3.3  \pm  0.5$\\
$<$4   &$ 7.9 \pm 1.1$ & $11.1  \pm  1.6$ &$5.1 \pm 0.7$ & $ 7.1  \pm  1.0$\\
$<$6   &$11.0 \pm 1.5$ & $17.4  \pm  2.4$ &$6.8 \pm 1.0$ & $10.8  \pm  1.5$\\
$<$9   &$14.2 \pm 2.0$ & $24.1  \pm  3.4$ &$8.9 \pm 1.3$ & $15.0  \pm  2.1$\\
\noalign{\smallskip}
\hline
\noalign{\medskip}
\end{tabular}}

\vskip3mm

\begin{minipage}{14cm}
{\eightrm NOTES --
Col. (1):  radius interior to which a given quantity is calculated, expressed in arc minutes.
Col. (2):  estimate of the cumulative mass based on the X-ray methodology, expressed in units of $10^{11}M_\odot$ for $T=0.80\pm 0.05$ keV.
Col. (3):  mass-to-light ratio based on the X-ray methodology, in the $B$-band, in solar units for $T=0.80\pm 0.05$ keV.
Col. (4):  estimate of the cumulative mass based on the TME methodology, expressed in units of $10^{11}M_\odot$ for the isotropic case.
Col. (5):  mass-to-light ratio based on the TME methodology, in the $B$-band, in solar units.
Col. (6):  estimate of the cumulative mass based on the MOND "simple" model (see text for details).
Col. (7):  mass-to-light ratio based on the MOND "simple" model, in the $B$-band, in solar units.
Col. (8):  estimate of the cumulative mass based on the MOND "standard" model (see text for details).
Col. (9):  mass-to-light ratio based on the MOND "standard" model, in the $B$-band, in solar units.
Col. (10):  estimate of the cumulative mass based on the MOND "toy" model (see text for details).
Col. (11):  mass-to-light ratio based on the MOND "toy" model, in the $B$-band, in solar units.
The distance $d=15.96$ Mpc was always used.}
\end{minipage}
\label{tab:mass_gc}

\leftline{\includegraphics[width=16.5cm]{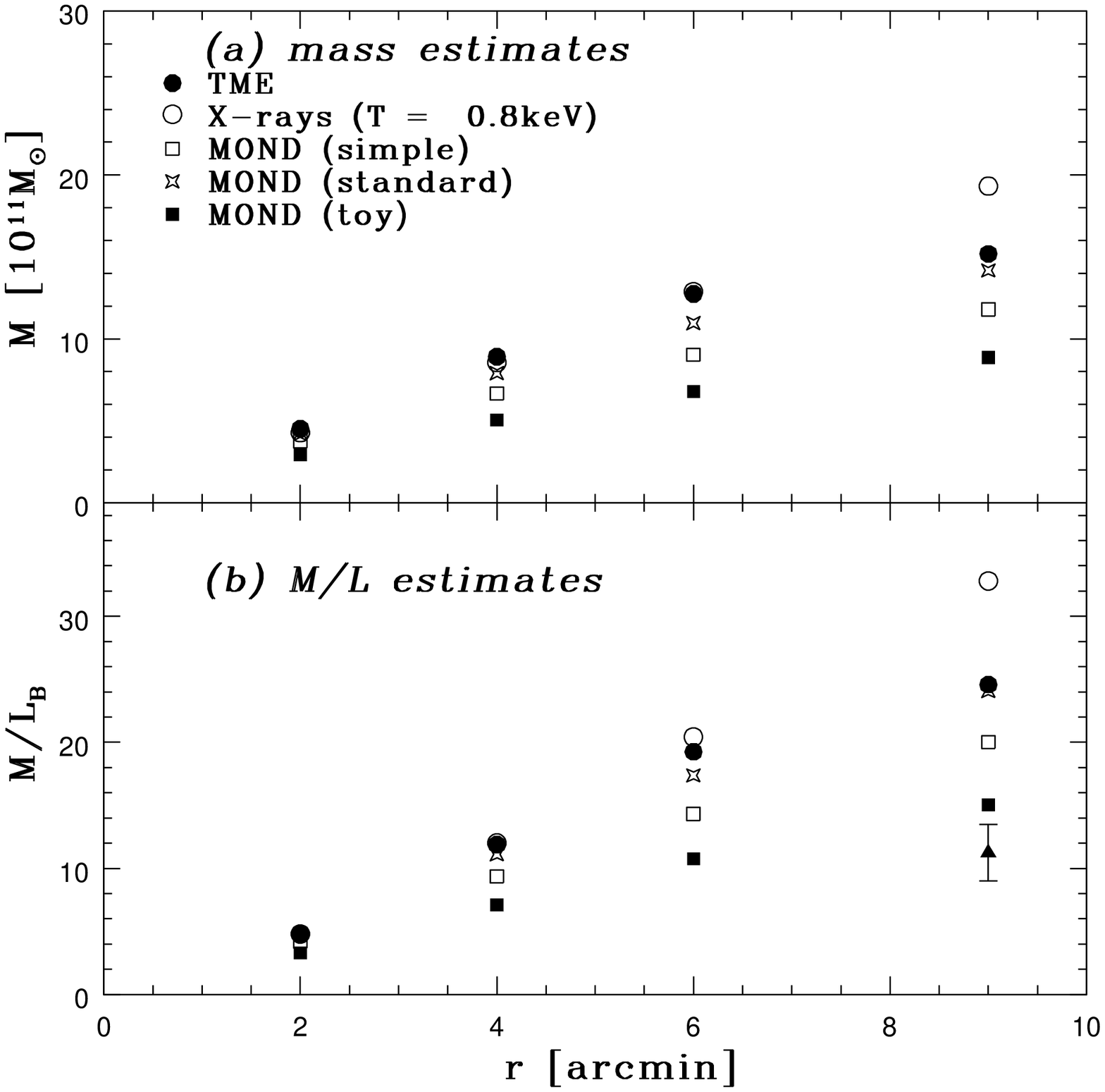}}
\figurecaption{1.}{\rm Total Newtonian and MOND mass (upper panel)
and Newtonian and MOND mass-to-light ratio (lower panel). In the
Newtonian approach we used "tracer mass estimates" and X-ray
halo with temperature $T=0.80$ keV according to the observational
data of Randall et al.~(2006). In the MOND approach we tested
three different models: "simple", "standard", and Bekenstein's
"toy" model (see text for details).  Additional point in the
lower panel at 9 arcmin given with the filled triangle
corresponds to the MOND estimate of the total mass-to-light ratio
based on the "toy" model and the TME methodology. The distance
$d=15.96$ Mpc is used everywhere.} 
\label{fig:NGC4649ML}

\begin{multicols}{2}

From Fig.~1 it can be seen that the estimates based on two physically very 
different methodologies are very similar (especially interior to $\sim 6$ 
arcmin), thus providing the strong evidence for the increase of mass (and the 
mass-to-light ratio) in NGC~4649 in Newtonian gravity. The discrepancy between 
the two methodologies and its consequences will be discussed below. 

\vfill\eject

\subsection{\bf 3.3 VIRIAL ESTIMATE}

In order to establish the total mass of  NGC~4649, we can
also use scalar virial theorem for a stationary stellar system
(Bertin et al.~2002):
\begin{equation}
{G\Upsilon_* L\over R_e}=K_V \sigma_0^2,
\label{eqn:virial1}
\end{equation}
where $\Upsilon_*$ is the {\it stellar} mass-to-light ratio in the given band, 
$L$ is the luminosity of the galaxy and $\sigma_0$ is the central velocity 
dispersion referred to an aperture radius of $R_e/8$. $K_V$ is the so-called 
"virial coefficient" which takes into account the projection effects. 
Cappellari et al.~ (2006) showed that $K_V=5\pm 0.1$ for a sample of 
early-type galaxies at redshift $z\sim 0$. From this equation the total 
dynamical mass is: 
\begin{equation}
M_{\hbox{\eightrm{dyn}}}=K_V {\sigma_0^2 R_e\over G}.
\label{eqn:dynmass}
\end{equation}
Using the aforementioned value for $K_V=5$, $\sigma_0=230$ \kms, and 
$R_e=6.97$ kpc, one obtains the total dynamical mass of NGC~4649: 
$M_{\hbox{\eightrm{dyn}}}=(4.2\pm 1.3)\times 10^{11}M_\odot$. This value takes 
into account only {\it random} motions and it does not take into account the 
rotational support so it is necessarily lower than the true value. Taking into 
account Eq.~(\ref{eqn:mrot}), we get at $3R_e$ 
$M_{\hbox{\eightrm{rot}}}(3R_e)= (0.9\pm 0.7)\times 10^{11}M_\odot$  (for 
$v_{\hbox{\eightrm{rot}}}=141$ \kms).  Thus, the total dynamical mass within 
$3R_e$ ($=3'.45$) is equal to $(5.1\pm 2.0)\times 10^{11}M_\odot$ which is 
very close to the values obtained using X-ray and TME methodologies at the 
same galactocentric distance (see Table 1). We note that in our estimate of 
the dynamical mass we have used the value of $\sigma_0$ inferred from 
the kinematics of the GCs. If we use the value of the velocity dispersion 
based on the integrated stellar spectra (Fisher et al.~1995), $\sigma_0=325$ 
\kms\ a higher value of the total dynamical mass (rotational support is 
assumed) is obtained: $M_{\hbox{\eightrm{dyn}}}=(9.3\pm 2.0)\times 
10^{11}M_\odot$. This is even closer to the values based on the TME and X-ray 
methodologies (see Table 1). 

The role of the rotation in early-type galaxies was discussed
recently by van der Wel et al.~(2008) who studied the sample of
these galaxies between redshift $z=1$ and the present epoch $z
\sim 0$ and found that for rotating galaxies $K_V$ is possibly
$\sim 20\%$ larger than the canonical value of 5. They used
Eq.~(9) only, in order to establish the total mass of the
galaxies in their sample. We see that in the case of NGC~4649, the
contribution of the rotation, equal to
$M_{\hbox{\eightrm{rot}}}/M_{\hbox{\eightrm{dyn}}}=0.9\times
10^{11}M_\odot /4.2\times 10^{11}M_\odot\approx 0.21$, is close
to that given by van der Wel et al., i.e. the mass is $\sim 21\%$
larger when we take the rotation into account, as given above.

\section{\bf 4.  TOTAL MASS IN MOND GRAVITY}

In this Section we calculate the total mass of NGC~4649 in MOND gravity using 
three different formulas: (i) the "simple" MOND formula from Famaey and 
Binney (2005), (ii) the "standard" formula (Sanders and McGaugh 2002) and 
(iii) the Bekenstein's "toy" model (Bekenstein 2004). The Newtonian 
acceleration is written as $a_N= a\mu(a/a_0)$ where $a$ is the MOND 
acceleration and $\mu (x)$ is the MOND interpolating function where 

\begin{equation}
x\equiv \frac{a}{a_0}.
\end{equation}
Here, $a_0=1.35^{+0.28}_{-0.42}\times
10^{-10}$ m~s$^{-2}$  is a new universal constant required by MOND (Famaey et
al.~2007). The interpolation function $\mu(a/a_0)$ shows the
asymptotic behavior, $\mu \approx 1$, for $a\gg a_0$, so that one
obtains the Newtonian relation in the strong field regime, and
$\mu=a/a_0$ for $a <$\hskip-1.2mm$ <a_0$.

As we have shown in Paper I the "external field effect" does not play an 
important role in the case of NGC~4649 so we will neglect it in our 
calculations below. 

The MOND dynamical mass, $M_M$ can be expressed by using the Newtonian one 
within a given radius $r$, $M_N$, through the following expression (e.g.~Angus 
et al.~2008): 

\begin{equation}
M_M(r)= M_N(r) \times \mu(x).
\label{eqn:mass_mond}
\end{equation}

The intepolation function can have different forms as given below (we refer 
the reader to Paper I regarding our MOND calculations and the
details related to the variable $x$ and the expressions for the circular 
velocity $v_{\hbox{\eightrm{circ}}}$ in MOND approach). We note that the 
Newtonian mass, $M_N$ is always based on Eq.~(\ref{eqn:TOT1}) unless stated 
otherwise. 

A "simple" MOND formula is given as:
\begin{equation}
\mu(x) = {x\over 1+x}.
\label{eqn:simple1}
\end{equation}
A "standard" MOND formula is given as:
\begin{equation}
\mu(x) = {x\over \sqrt{1+x^2}}.
\label{eqn:std1}
\end{equation}

Finally, for the "toy" model of Bekenstein the MOND formula is:

\begin{equation}
\mu(x) = {-1+\sqrt{1+4x}\over  {1+\sqrt{1+4x}}}.
\label{eqn:toy1}
\end{equation}

The results for the total mass and mass-to-light ratio are given
in Fig.~1 and Table 1. The
uncertainties due to the  error for the temperature and
the errors for the $a_0$ parameter are also given in Table
1.   All predictions imply the existence of dark
matter beyond $\sim 3R_e$, with the exception of  the Bekenstein
"toy" model based on the TME methodology, which is marginally
consistent with no dark matter hypothesis (see
Fig.~1).

\section{\bf 5. DISCUSSION AND CONCLUSIONS}

We used the kinematics of 121 GCs and the X-ray observations of
the early-type galaxy NGC~4649 in order to examine its mass
profile. We performed our calculations in  Newtonian and MOND
approaches with the goal of establishing the existence (or lack)
of dark matter in this galaxy. In Paper I we found that the
velocity dispersion of this galaxy can be fitted with $M/L_B\sim
7$ implying a low amount of dark matter, if any. Judging from the
approximately constant value of the velocity dispersion in the bins at
various galactrocentric distance from the center we found a hint
of increasing mass-to-light ratio in the outer bins, i.e. we could
not exclude the mass-to-light ratio as high as 15 in the
outermost bin (beyond $\sim 8$ arcmin) (see Fig. 3 of Paper I).

Our results are as follows:

1) Interior to $\sim 3R_e$ ($=$270 arcsec) we find no support for the 
significant amounts of dark matter. Both TME and X-ray methodologies predict 
the mass-to-light ratio $M/L_B\sim 12$ (in the Newtonian approach) which is 
higher than the value obtained using the Jeans modelling (see Paper I) but 
given the error bars (see Table 1) we see that the amount of dark matter may 
indeed be very low. In the MOND approach the predicted mass-to-light ratio is 
lower (between $\sim 7$ and $\sim 9$) which is obviously consistent with the 
absence of dark matter interior to this radius. The virial estimate of the 
total dynamical mass based on the velocity dispersion inferred from the 
kinematics of the GCs ($\sigma_0=230$ \kms) is lower than the estimate which 
we obtain using the velocity dispersion based on the integrated stellar 
spectra ($\sigma_0=325$ \kms); the latter value is in a very good agreement 
with the estimates obtained using both X-ray and TME methodologies. 

2) Beyond $\sim 3R_e$ the situation becomes more complex: both
X-ray and TME methodologies predict rapid increase of the total mass
(and the mass-to-light ratio) implying the existence of the dark
component which starts to dominate. Both methodologies provide a
very similar prediction: interior to 6 arcmin ($4R_e$) the
mass-to-light ratio becomes $M/L_B\sim 20$ which implies at least
50\% of the dark matter contribution. At 9 arcmin ($6R_e$)   the
predictions of the two methodologies differ: the X-rays estimate
($M/L_B\sim 33$ is much higher than that based on TME, $M/L_B\sim
25$): note that the estimates are nevertheless consistent given the larger
error bars in the last bin. The predictions of both methodologies
imply significant amounts of dark matter in the outer regions of
NGC~4649 in the Newtonian approach. MOND estimates are lower but
for the estimates based on the X-rays one still needs significant
amounts od dark matter in the outer regions. There is only one
case for which we managed to have rather low mass-to-light ratio
in the MOND approach: this is a case based on the TME methodology
and the Bekenstein's "toy" formula. At 9 arcmin $M/L_B=11.3\pm
2.2$ which is marginally consistent with the absence of dark
matter (see Fig. 1). While this marginal consistency may
present the "last resort" for the MOND proponents, one should
also point out that all other encouraging results of the MOND
approach have been obtained using different models (which we have
dubbed "simple" and "standard"), which in this case require large
quantities of dark matter, thus obviating the motivation for such
a radical approach as MOND in the first place.

3) The question of the discrepancy between the X-ray  and TME mass estimates 
remains open. One of the possible reasons could be the deviation from 
hydrostatic equilibrium in NGC~4649. We do not consider this as a crucial 
issue in the present case (see the discussion at the end of Section 3.1). One 
possible reason can be the lower gas temperature in the outer regions of 
NGC~4649. This could be related to the reheating in the central regions, along 
the lines set by the model of evolving cooling flows by Binney and Tabor 
(1995). Note that the last measured point in the study of Randal et al.~(2006, 
table 4) is interior to $\sim 200$ arcsec which is well inside $3R_e$, i.e. 
within the region where all approaches agree on the predicted total mass. In 
order to resolve this question it is necessary to have the X-ray observations 
which extend beyond $\sim 3R_e$ and to have more observed GCs in the outer 
bins so as to reduce the uncertainties related to the kinematical profile of 
NGC~4649. 

\acknowledgements{This work was supported by the Ministry of Science of the 
Republic of Serbia through the project no.~146012, "Gaseous and stellar 
component of galaxies: interaction and evolution". This research has made use 
of the NASA/IPAC Extragalactic Database (NED) which is operated by the Jet 
Propulsion Laboratory, California Institute of Technology, under contract with 
the National Aeronautics and Space Administration and NASA Astrophysics Data 
System.  We acknowledge the usage of the HyperLeda database 
(http://leda.univ-lyon1.fr). We thank our referee, S. Ninkovi\'c, for a 
detailed and very useful report which improved the quality of the manuscript.}

\vskip.5cm

\references

\vskip-2mm

Angus, G. W., Famaey, B. and Buote, D. A.: 2008, \journal{Mon. Not. R. Astron. 
Soc.}, \vol{387}, 1470 

Bekenstein, J.: 2004, \journal{Phys. Rev. D}, \vol{70}, 083509.

Bertin, G., Ciotti, L. and Del Principe, M.: 2002, \journal{Astron. 
Astrophys.}, \vol{386}, 149. 

Binney, J. and Tabor, G.: 1995, \journal{Mon. Not. R. Astron. Soc.}, \vol{276}, 
663. 

Brown, B. A. and Bregman, J. N.: 2001, \journal{Astrophys. J.}, \vol{547}, 
154. 

Cappellari, M., Bacon, R., Bureau, M. et al.: 2006, \journal{Mon. Not. R. 
Astron. Soc.}, \vol{360}, 1126.

de Vaucouleurs, G., de Vaucouleurs, A. Corwin, H. G. Jr. et al.: 1991, Third 
Reference Catalogue of Bright Galaxies, Springer-Verlag, New York. 

Diehl, S. and Statler, T. S.: 2007, \journal{Astrophys. J.}, \vol{668}, 150.

Evans N. W., Wilkinson M. I., Perrett K. M., Bridges T. J.: 2003, 
\journal{Astrophys. J.}, \vol{583}, 752. 

Famaey, B. and Binney, J.: 2005, \journal{Mon. Not. R. Astron. Soc.}, 
\vol{363}, 603. 

Famaey, B., Gentile, G., Bruneton, J.-P. and Zhao, H. S.: 2007a, 
\journal{Phys. Rev. D}, \vol{75}, 063002  

Fisher, D, Illingworth, G. and Franx, M.: 1995, \journal{Astrophys. J.}, 
\vol{438}, 539. 

Kim, D.--W. and Fabbiano, G.: 1995, \journal{Astrophys. J.}, \vol{441}, 182.

Kim, E., Kim, D.--W., Fabbiano, G. et al.: 2006, \journal{Astrophys. J.}, 
\vol{647}, 276. 

Komatsu, E., Dunkley, J., Nolta, M. R. et al.: 2008, \journal{Astrophys. 
J. SS}, submitted {\tt [arXiv: astro-ph/0803.0547]} 

Lee, M. G., Hwang, H. S, Park, H. S. et al.: 2008, \journal{Astrophys. J.}, 
\vol{674}, 857. 

Milgrom, M.: 1983, \journal{Astrophys. J.}, \vol{270}, 365.

Randall, S. W., Sarazin, C. L. and Irwin, J. A.: 2006, \journal{Astrophys. 
J.}, \vol{636}, 200. 

Samurovi{\'c}, S.: 2006, \journal{Serb. Astron. J.}, \vol{173}, 35.

Samurovi{\'c}, S.: 2007, Dark Matter in Elliptical Galaxies, \journal{Publ. 
Astron. Obs. Belgrade}, {\bf 81}. 

Samurovi{\'c}, S. and Danziger, I. J.: 2005, \journal{Mon. Not. R. Astron. 
Soc.}, \vol{363}, 769. 

Samurovi{\'c}, S. and Danziger, I. J.: 2006,  \journal{Astron. Astrophys.}, 
\vol{458}, 79. 

Samurovi{\'c}, S. and \'Cirkovi\'c, M. M.: 2008, \journal{Astron. Astrophys.}, 
\vol{488}, 873 

Sanders, R. H. and McGaugh, S.: 2002, AR \journal{Astron. Astrophys.}, 
\vol{40}, 263. 

Tegmark, M. et al.: 2004, \journal{Phys. Rev. D}, \vol{69}, 103501.

van der Marel, R. P.: 1991, \journal{Mon. Not. R. Astron. Soc.}, \vol{253}, 
710.

van der Wel, A., Holden, B. P., Zirm, A. W., Franx, M., Rettura, A., 
Illingworth, G. D. and Ford, H. C.: 2008, \journal{Astrophys. J.}, accepted, 
{\tt [arXiv: astro-ph/0808.0077v1]} 

\endreferences 

\end{multicols}

\end{document}